\documentclass[12pt]{article}
\usepackage{graphicx}

\newcommand\pubnumber{CIPANP2015-EW-Bodek, CMS CR-2015/201}
\newcommand\pubdate{\today}

\def\rochester{for the  CDF, D0, CMS and ATLAS Collaborations\\
Department of Physics and Astronomy\\ University of Rochester, Rochester, NY 14627, USA}
\def\support{\footnote{Work supported by the US. Department of Energy.}}

\def\Title#1{\begin{center} {\Large #1 } \end{center}}
\def\Author#1{\begin{center}{ \sc #1} \end{center}}
\def\Address#1{\begin{center}{ \it #1} \end{center}}

\newcommand\pubblock{\rightline{\begin{tabular}{l} \pubnumber\\
         \pubdate  \end{tabular}}}
\newenvironment{Abstract}{\begin{quotation}  }{\end{quotation}}
\newenvironment{Presented}{\begin{quotation} \begin{center} 
             PRESENTED AT\end{center}\bigskip 
      \begin{center}\begin{large}}{\end{large}\end{center} \end{quotation}}

\def\beq{\begin{equation}}
\def\eeq#1{\label{#1}\end{equation}}
\def\eeqn{\end{equation}}

\def\beqa{\begin{eqnarray}}
\def\eeqa#1{\label{#1}\end{eqnarray}}
\def\eeqan{\end{eqnarray}}

\let\bar=\overbar

\def\Dslash{\not{\hbox{\kern-4pt $D$}}}
\def\dslash{\not{\hbox{\kern-2pt $\del$}}}

\def\msb{{\bar{\ssstyle M \kern -1pt S}}}

\begin{document}
\begin{titlepage}
\pubblock

\vfill
\Title{Standard Model Precision Electroweak Measurements at HL-LHC and Future Hadron Colliders }
\vfill
\Author{ A.  Bodek\support}
\Address{\rochester}
\vfill
\begin{Abstract}
We investigate the uncertainties  for current and  future measurements of  electroweak (EW) parameters at hadron colliders.  These include  the measurement of the mass of the  top quark ($M_T$), the direct measurement of the mass of the  W boson ($M_W^{direct}$), the measurement of the effective EW mixing angle   $\sin^2\theta^{\rm lept}_{\rm eff}$(M$_Z$), and the  measurement of the on-shell EW mixing angle  $\sin^2\theta_{W}= 1- M_W^2/M_Z^2$ which is equivalent to an indirect measurement of the W mass ($M_W^{indirect}$). Reduction of a factor of 2 to 3 in the measurement  errors is expected in the future.
\end{Abstract}
\vfill
\begin{Presented}
The 12th Conference on the Intersection of Nuclear and Particle Physics, 
CIPANP 2015\\  May 19-24, 2015 (Vail, CO, USA) 
\end{Presented}
\vfill
\end{titlepage}
\def\thefootnote{\fnsymbol{footnote}}
\setcounter{footnote}{0}
\section {Measurements of electroweak parameters.}
With the discovery of the Higgs boson, the standard model is over constrained. Within the standard model (SM), measurements of the mass of the $Z$ boson ($M_Z$) and the mass of the  top quark ($M_T$), in combination with the
    mass of the Higgs boson ($M_H$),  can be used to predict  the mass of the W boson ($M_W$).
  
   \begin{figure}[h]
 \centering
\includegraphics[width=6.in,height=3.5in]{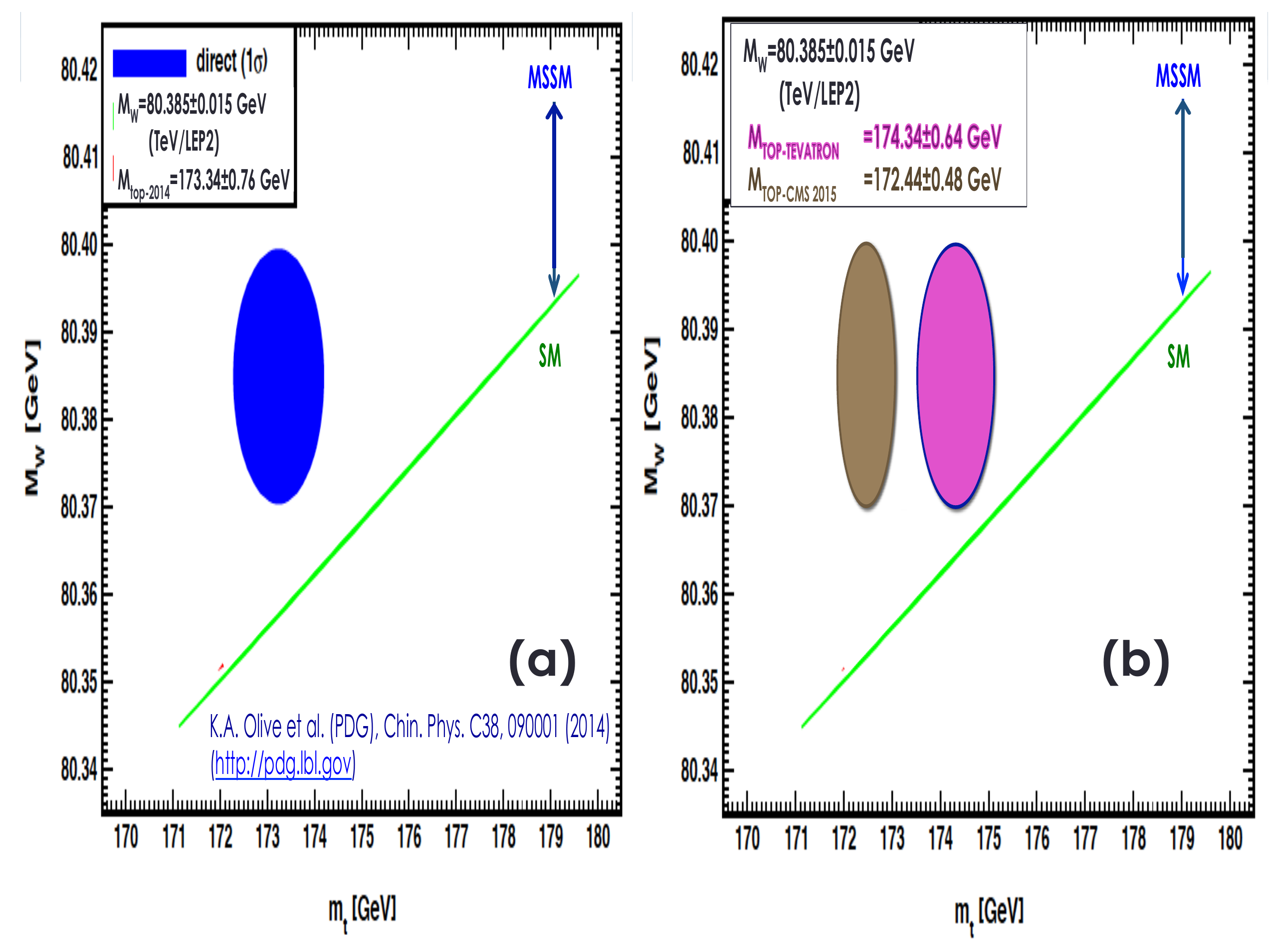}
\caption{ (a) World average of all direct measurements of $M_W$  (CDF, D0, LEP2) versus 
the average of all  $M_T$ measurements  (CDF, D0, CMS, ATLAS) in  2014.  Also shown is the expectation from the SM (with $M_H=125.6\pm0.7$ GeV) in green.  Supersymmetry models predict values which are above the SM line.  (b) Same as (a) but with the CMS measurement of  $M_T$ in 2015 as compared to the Tevatron
measurement of $M_T$.
 }
\label{Fig1}
\end{figure}
  Fig.\ref{Fig1} (a)  (from ref.\cite{PDG}) shows the current world average\cite{w2014}  of direct measurements of $M_W$=$80.385\pm0.015$ GeV versus the average\cite{top2014} of the direct measurements of  $M_T$=
$173.34\pm0.76$ GeV.  Also shown in green is the expectation from the SM with $M_H=125.6\pm0.7$ GeV.   The average of  all direct measurements of $M_W$
    is about 1.5 standard deviation higher  than the prediction of the standard
    model.  Predictions of  supersymmetric models
     for $M_W$ are also higher \cite{super} than the predictions of the standard model.
     Therefore, more precise measurements of $M_W$
    are of great  interest.
    
    Alternatively, $M_W$  can also be extracted indirectly from
     measurements of the on-shell electroweak mixing angle $\sin^2\theta_{W}$ 
    by  the  relation $\sin^2\theta_{W}=1-M_W^2/M_Z^2$.
In this communication we investigate the uncertainties  in  future measurements of  electroweak (EW) parameters at hadron colliders.  These measurements include  $M_W$, $M_T$,  the effective EW mixing angle 
$ \sin^2\theta^{\rm lept}_{\rm eff}$(M$_Z$), and the on-shell EW mixing angle  $\sin^2\theta_{W}= 1- M_W^2/M_Z^2$ which is equivalent to an indirect measurement of $M_W$.
\subsection{Direct measurements of $M_T$}
  The average of the Tevatron measurements of $M_T$ in 2014 is 174.34 $\pm$ 0.37 $\pm$ 0.52 GeV (174.34$\pm$0.64).  When combined with  the 2014 measurements of ATLAS and CMS in March 2014, the combined  2014 world average \cite{top2014}  (CDF, D0, CMS, ATLAS) is 173.34 $\pm$ 0.27 $\pm$ 0.71 GeV (or 173.34 $\pm$ 0.76 GeV). 
  
   \begin{figure}[h]
 \centering
\includegraphics[width=6.0in,height=3.2in]{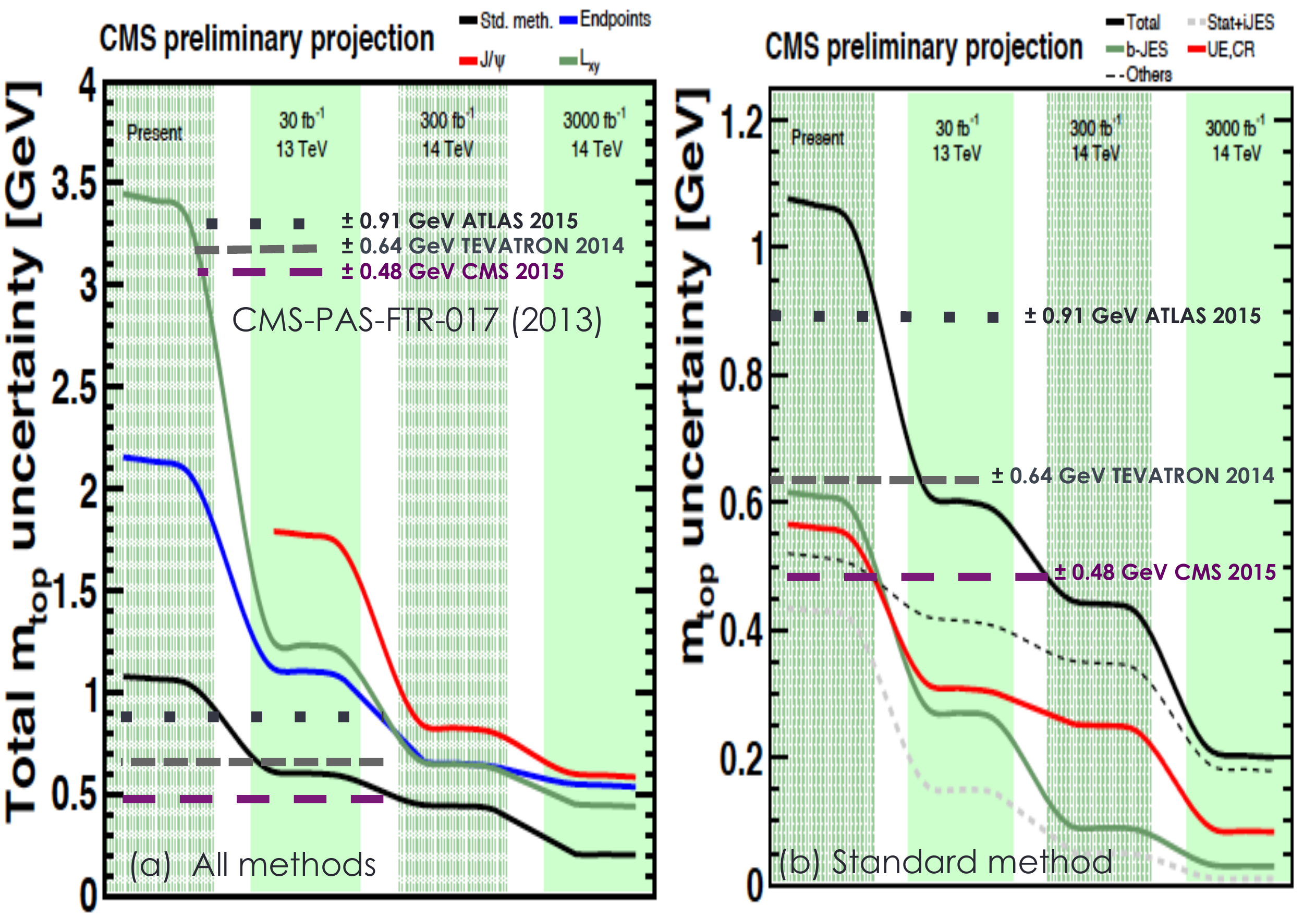}
\caption{ (a) Projection of the uncertainty in the measurement of $M_T$ (in GeV) at CMS using
different methods for various integrated luminosities. (b) Various contributions to the error in $M_T$ (in GeV)  using the standard method for the measurement.  Figures are from ref.~\cite{CMS_M_T_future}.}
\label{Fig2}
\end{figure}
  The most recent measurement  of $M_T$ at the LHC are somewhat lower than at the Tevatron. 
  The  ATLAS result\cite{ATLAStop} published  2015 is $M_T$= 172.99 $\pm$ 0.91 GeV.   The CMS\cite{CMStop} 2015 measurement of  $M_T$=172.44 $\pm$ 0.13 $\pm$ 0.47 GeV (172.44 $\pm$0.48 GeV) is the most precise measurement to date  and supersedes all previous CMS results.  There is about a 2 standard deviation tension between CMS measurement of $M_T$ in 2015 and  the earlier Tevatron measurements.  However,  both are consistent with the world average.   As shown in Fig.~\ref{Fig1} (b) , a lower value of $M_T$ would imply a somewhat larger deviation of $M_W$ from the prediction of the SM. 
   
As of  2015 CMS achieved\cite{CMStop}  a $\pm$0.48 GeV uncertainty in the measurement of $M_T$. Fig.~\ref{Fig2}(a) shows the projected uncertainty in the measurement of $M_T$ (in GeV) at CMS using different methods for various integrated luminosities.  Fig.~\ref{Fig2}(b) shows the various contributions to the error  in $M_T$ (in GeV) using the standard method for the measurement. 
 In the long term,  as shown in Fig.~\ref{Fig2},  a precision of  $\pm$0.2 GeV could be achieved\cite {CMS_M_T_future} with about 3000 fb$^{-1}$ of data at 14 TeV. 
 From the experimental point of view it is conceivable that an uncertainty of $O(\Lambda_{QCD})$ is achieved at the end of HL-LHC.  
In the sub-GeV regime issues of theoretical interpretation become important,
in particular the understanding of the relation of the top quark pole mass with that
deployed in the simulations used for the calibration of  the measurement of $M_T$ in hadron colliders.
 
  %

\subsection{Direct measurements of $M_W$}
Both  D0 and CDF  published direct measurements\cite{w2014} of  $M_W$ using data
samples of the first 2.2 $fb^{-1}$ of run II at the Tevatron.  The uncertainty in the results from
each experiment is about 20 MeV.  The combined results of the two experiments has an
error of 15 MeV.   The sources
of the uncertainty in the CDF\cite{CDF_MW} published  2.2 $fb^{-1}$ result are listed in Table~\ref{Table1}.

Analysis of the full  9.4$fb^{-1}$ Teatron run II sample is currently under way. One would expect a reduction
in the statistical error of about a factor of 2 (from 12 to 6 MeV).  The dominant systematic errors from
the energy scale and Parton Distribution Functions (PDFs) could be reduced significantly as discussed later in this
paper.   Therefore, a  10 MeV uncertainty in the direct measurement of  the W mass  may be achievable with the full run II data sample at the Tevatron.

\begin{table}[t]
\begin{center}
\begin{tabular}{|l|c|} 
\hline\hline
Source  of uncertainty in CDF measurement  of $M_W$ &   Uncertainty  \\  
with 2.2  fb$^{-1}$  from Phys. Rev. Lett. 108, 151803 (2012) & (MeV) \\ \hline\hline
Lepton energy scale and resolution    &    7 \\
Recoil energy scale and resolution    &    6 \\
Lepton removal    &    2 \\
Backgrounds    &    3 \\
$p_T$(W)    &   5 \\
Parton distributions (PDFs)    &    10 \\
QED radiation    &    4 \\
\hline 
W-bodon statistics   &    12 \\
 \hline \hline
Total    &    19 \\
 \hline\hline
\end{tabular}
\caption{Sources of uncertainty in the CDF 2.2. fb$^{-1}$ direct measurement of $M_W$ at the Tevatron. Table from ref. \cite{CDF_MW} (Phys. Rev. Lett. 108, 151803 (2012)).}
\label{Table1}
\end{center}
\end{table}

The  current ongoing measurement of $M_W$ at CMS is  using the lower luminosity (lower pileup) sample
 taken at 7 TeV (with $\approx 5~fb^{-1}$).   The current  ongoing LHC analyses also aim at an error of
about 10 MeV.  Because of the sensitivity of the measurement 
of missing $E_T$ to the effects of  pileup,  the higher integrated luminosities at 8~TeV and at 
13-14 TeV may not be as useful for the direct measurement of  $M_W$. 
However,  as discussed in following sections of this paper,
data with higher luminosities can contribute to constraining PDFs and thus help reduce the PDF
error in the measurement of $M_W$ with  existing low pileup LHC data samples.
\begin{figure}[h]
 \centering
\includegraphics[width=6.0in,height=4.2in]{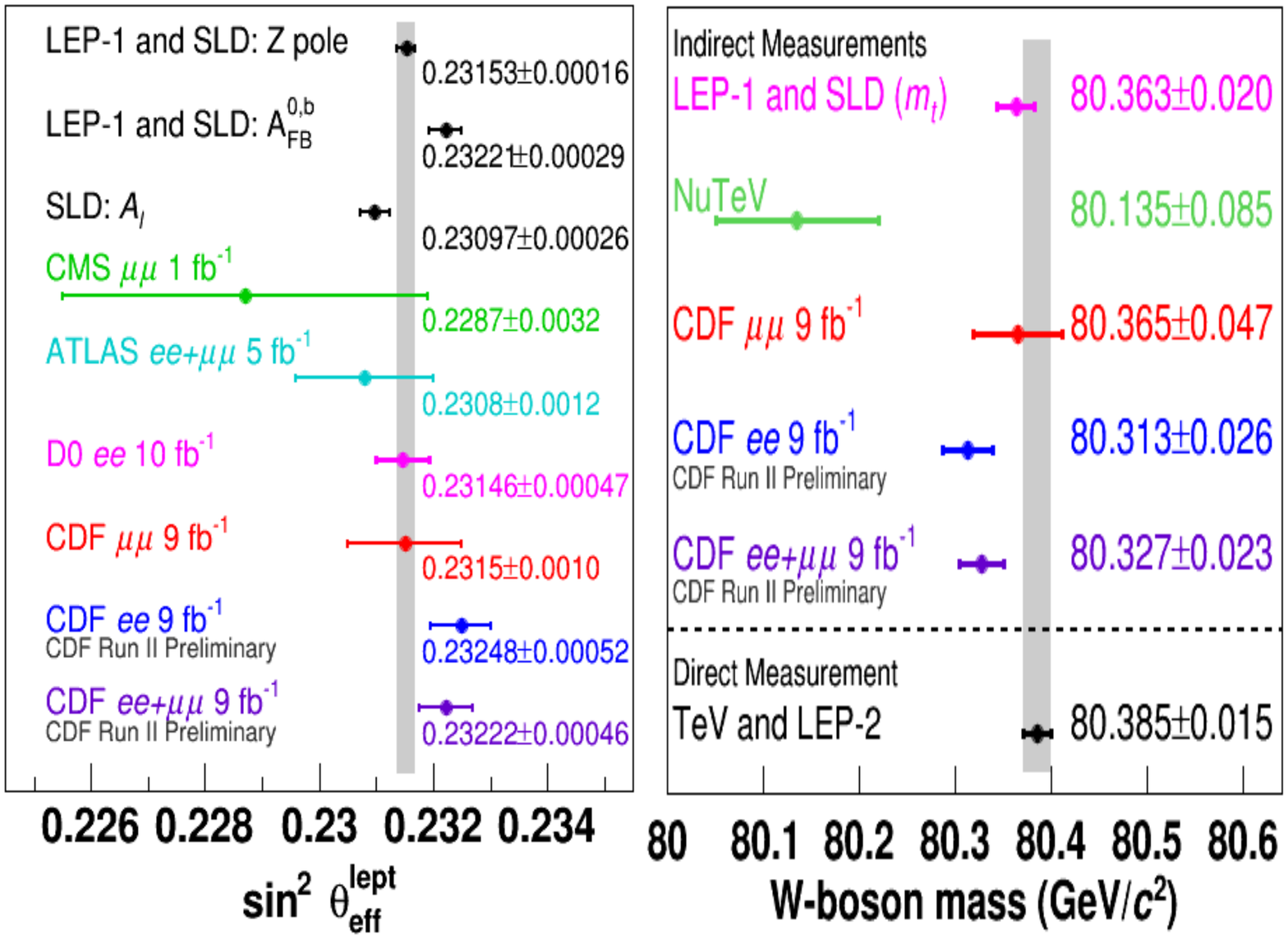}
\caption{ (a) Word measurements of $\sin^2\theta_W$. (b) Word measurements $M_W^{direct}$ and  $M_W^{indirect}$.  Note that the value from LEP-1/SLD is an average of six measurements. Figures are from ref.~\cite{CDFweb}.}
\label{Fig3}
\end{figure}
\subsection{Tevatron measurements of $\sin^2\theta_W$ and  $M_W^{indirect}$}
Measurements of the  forward-backward  charge asymmetry ($A_{FB}(M,y)$  in Drell-Yan
         dilepton events produced at hadron colliders 
(in the region of the $Z$ pole)   have been used to measure
 the value of the  {\it effective} electroweak (EW)  mixing
  angle   $\sin^2\theta^{\rm lept}_{\rm eff}$(M$_Z$)\cite {cdf-ee,cdf-mumu,Dzero,CDFweb,ATLAS,LHCb}.
   In addition, by incorporating electroweak radiative corrections in the analysis, 
   the CDF collaboration has also extracted  the  {\it on-shell} EW mixing angle  $\sin^2\theta_W =1- M_W^2/M_Z^2$~\cite{cdf-ee,cdf-mumu,CDFweb}.
  
   An error of $\pm$0.00030 in the measurement of $\sin^2\theta_{W}$ 
      is equivalent to an indirect measurement  of the W mass to a precision of $\pm$15 MeV. 
     Within the SM,  the direct and indirect measurements of $M_W$ should agree with  each other.
Since the standard model is over constrained,  any inconsistency between precise measurements  of SM parameters would be indicative of new physics.   

Similarly, in order to help resolve the  long standing  $3.2\sigma$ discrepancy\cite{lep-sld} 
 between the two most precise LEP/SLD Z pole measurements of  $\sin^2\theta^{\rm lept}_{\rm eff}(M_Z)$,  new measurements of   $\sin^2\theta^{\rm lept}_{\rm eff}(M_Z)$ should have errors similar to SLD or  LEP ($\approx \pm$0.00030).
 \begin{eqnarray}
 \sin^2\theta^{\rm lept}_{\rm eff}(LEP/SLD~ A_{FB}^{0,b})&=&0.23222\pm0.00029\\
 \sin^2\theta^{\rm lept}_{\rm eff}(SLD ~A_{LR})&=&0.23098\pm0.00026
  \end{eqnarray}

\begin{table}[h]
\begin{center}
\begin{tabular}{|l|c|c||} 
\hline\hline
Source uncertainty in CDF measurement  &   Uncertainty  in &  Uncertainty  in \\  
with 9.4  fb$^{-1}$ (electron+muons)& $\sin^2\theta^{\rm lept}_{\rm eff}$(M$_Z$) & $M_W^{indirect}$ (GeV) \\ \hline\hline
Data: Statistics    &  $\pm$0.00042 (stat)  & 0.020\\ \hline
Data:  Energy scale   &   $\pm$0.00003 (syst) &0.001 \\
Data:  Backgrounds   &   $\pm$0.00002 (syst)  & 0.001 \\\hline
Prediction:  PDFs  &    $\pm$0.00016 (syst)  &0.008\\
Prediction: QCD EBA (NLO minus LO)   &   $\pm$0.00007 (syst)  &  0.003\\
Prediction: QCD scales   &    $\pm$0.00002 (syst) &  0.001 \\ \hline
All systematics  &   $\pm$0.00018 (syst)   & 0.009 \\
 \hline \hline
Total: (stat+syst)     &  $\pm$0.00046(total)    & 0.023 \\
 \hline\hline
\end{tabular}
\caption{Sources of uncertainty in the CDF 9.4  fb$^{-1}$  measurements of $\sin^2\theta^{\rm lept}_{\rm eff}$(M$_Z$)  and  $M_W^{indirect}$ at the Tevatron (ref.~\cite{CDFweb}).}
\label{Table2}
\end{center}
\end{table}

More precise extractions of $\sin^2\theta^{\rm lept}_{\rm eff}$and the on-shell  $\sin^2\theta_W= 1 - M_W^2/M_Z^2$  using $A_{\rm fb}(M,y)$ measurement for  dilepton events produced in 
p$\bar p$ and pp collisions  are now  possible  because of  the following four novel techniques:

\begin{itemize}
\item A new technique \cite{scale} for calibrating muon  and electron energy scales
as a function of  detector  $\eta$ and $\phi$ (and sign), thus greatly reducing systematic            
errors from the energy scale.   This technique is used in CDF, D0 and CMS.

\item A new  event angle weighting technique\cite{weighting} for the measurement of $A_{FB}$. With this technique
all  experimental uncertainties in acceptance and efficiencies cancel (by measuring  the $\cos\theta$ coefficient $A_4$  and  using the relation  $A_{FB}=(8/3)A_4$). Similarly,  additional weights can be  included for antiquark dilution, which makes the analysis independent of the acceptance in dilepton rapidity.   This technique is used in CDF
\cite{cdf-ee,cdf-mumu, CDFweb} and is currently being implemented at CMS.

\item The  implementation\cite{cdf-ee} of Z fitter Effective Born Approximation (EBA) electroweak radiative corrections into the theory predictions of POWHEG\cite{powheg}  and RESBOS\cite{resbos}   which 
 allows for a measurement of both  $\sin^2\theta^{\rm lept}_{\rm eff}$(M$_Z$) and 
  $\sin^2\theta_W= 1 - M_W^2/M_Z^2$.   These EBA  electroweak radiative corrections were implemented in  CDF analyses\cite{cdf-ee,cdf-mumu, CDFweb} since  2013.
  Recently, a  POWHEG version with electroweak radiative corrections has been released.  Similarly, electroweak radiative corrections have been implemented in other theory predictions. Comparisons of different  implementation of EW radiative corrections  are now possible.
  
  \item  The use of Drell-Yan  $A_{FB}(M,y)$  ($\chi^2$ weighting) first proposed in ref. \cite{pdf_error}) for additional constraints on PDFs.  The $\chi^2$ weighting  technique reduces the PDF error in the   measurements of 
  $\sin^2\theta^{\rm lept}_{\rm eff}$(M$_Z$),  $\sin^2\theta_W$, and  in the indirect and
  direct measurements of   $M_W$.  This technique is now used in CDF\cite{CDFweb} and is currently being implemented in CMS.
 \end{itemize} 
Fig.~\ref{Fig3}(a) from ref.~\cite{CDFweb} shows the CDF measurements of   $\sin^2\theta^{\rm lept}_{\rm eff}$(M$_Z$) (extracted from the full run II 9.4 fm$^{-1}$ sample)
compared to other measurements. Fig.~\ref{Fig3} (b)  shows the corresponding CDF indirect measurements  $M_W$  ($M_W^{indirect}$)  compared to other indirect and  direct measurements of $M_W$.  .

Table~\ref{Table2} lists the sources of uncertainty in the CDF 9.4  fb$^{-1}$  measurements~\cite{CDFweb} of
 $\sin^2\theta^{\rm lept}_{\rm eff}$(M$_Z$)  and  $M_W^{indirect}$ at the Tevatron.
 The 23  MeV error  in the $indirect$ measurement of $M_W$ at CDF extracted from full  9.4 fm$^{-1}$ data  sample
(shown in Table\ref{Table2}  is similar to the 19 MeV  error in the $direct$ measurement of $M_W$ extracted from  2.2  fm$^{-1}$ sample (shown in Table~\ref{Table1}).  
  The use of Drell-Yan  $A_{FB}$ for additional constraints ($\chi^2$ weighting) 
  on PDFs \cite{pdf_error} reduces the NNPDF 3.0 (NNLO) PDF error  on the CDF measurement of 
 $\sin^2\theta^{\rm lept}_{\rm eff}$(M$_Z$) from $\pm$ 0.00020 to
   $\pm$ 0.00016 as described below.   
   
   The errors in the CDF and D0
 measurements of $\sin^2\theta^{\rm lept}_{\rm eff}$(M$_Z$) are $\approx \pm0.00046$.  An official combination of the CDF ($e^+e^+\mu\mu$) and D0 ($e^+e^-$) results has not yet been done, but the error in the combination of  $\sin^2\theta^{\rm lept}_{\rm eff}$(M$_Z$) 
 would be about $\pm0.00038$ (which is equivalent to an error of 19 MeV in $M_W^{indirect}$). The error in the combination would be smaller when the analysis of the D0 $\mu\mu$ sample is completed

\subsubsection {Constraining PDFs through $\chi^2$ weighting}
 This technique which  was  first proposed in ref. \cite{pdf_error} has  been implemented in the most recent CDF analysis\cite{CDFweb}. At the Tevatron the technique 
 reduces the PDF error by 20\%.    The reduction of the PDF error at the LHC is much more significant.
 
\begin{figure}[h]
 \centering
\includegraphics[width=6.in,height=3.9in]{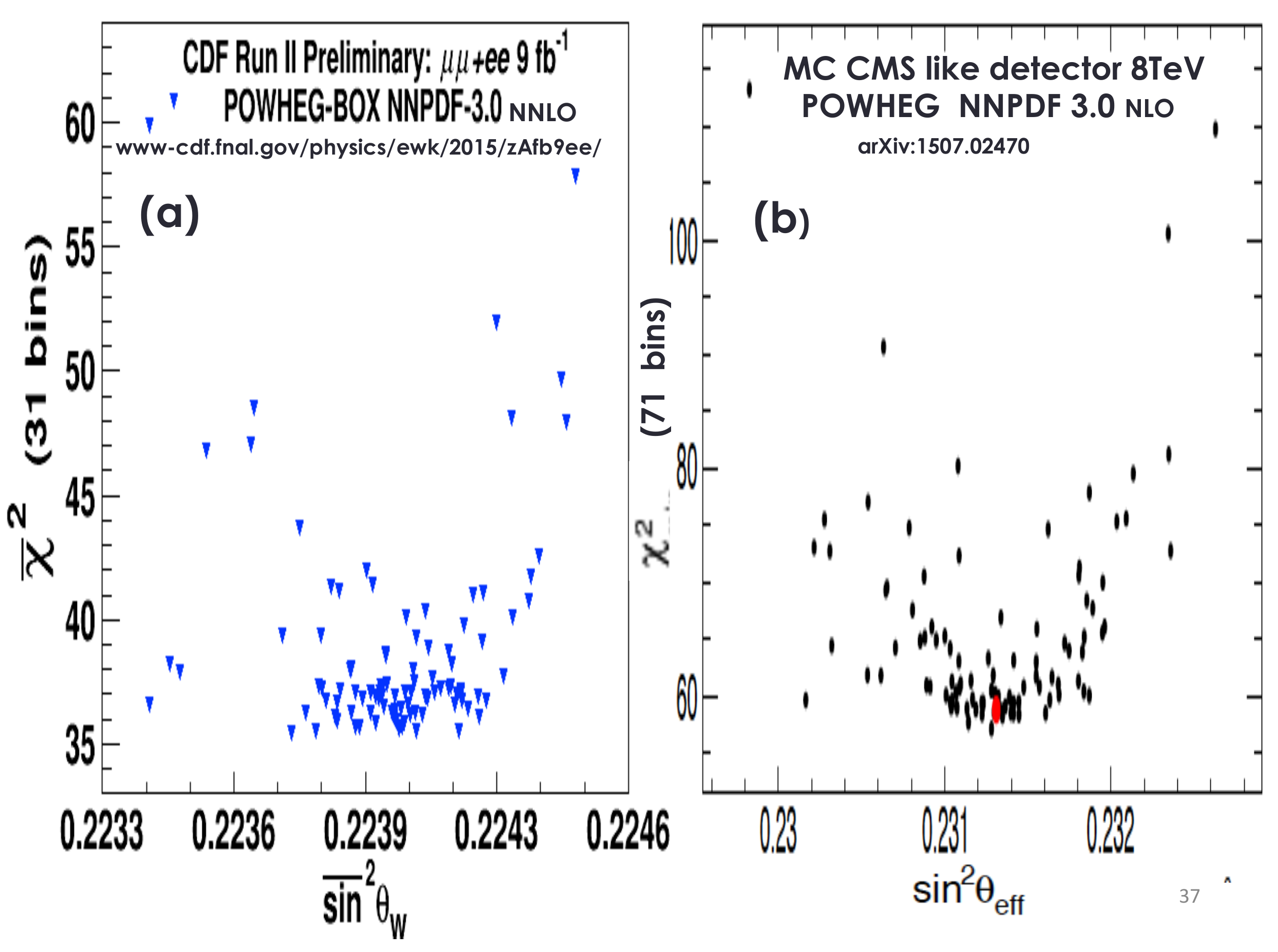}
\caption{ (a) CDF data: Best $\chi^2$ versus $\sin^2\theta_W$ (from ref. \cite{CDFweb}). (b) 
Best $\chi^2$ versus   $\sin^2\theta^{\rm lept}_{\rm eff}$(M$_Z$)  for MC simulation of a  CMS like detector with  15 fb$^{-1}$ at 8 TeV  (from ref.~\cite{pdf_error} (arXiv:1507.02470).}
\label{Fig4}
\end{figure}

 Fig.~\ref{Fig4}(a) shows the $\chi^2$ for the best fit value of $\sin^2\theta_{W}$  at CDF for each of the 100 PDF replicas for the NNPDF 3.0 (NNLO) PDF set.  As shown in ref.\cite{pdf_error} different values of $\sin^2\theta_{W}$ raise or lower $A_{FB}$(M) for all values of  dilepton mass.  In contrast, PDFs which raise the value of $A_{FB}$ for dilepton mass above the mass of the Z boson, reduce $A_{FB}$ below the mass of the Z bosons.  The sensitivity
 of $A_{FB}(M)$ to $\sin^2\theta_{W}$ is very different from the sensitivity to PDFs. 
 Therefore, PDFs with a high 
 value of $\chi^2$ are less likely to be correct.  
 As shown in ref. \cite{pdf_error}, this information can be incorporated into the analysis by weighting the PDF replicas by $e^{-\chi^2/2}$. This greatly reduces the weights of PDFs with large values of   $\chi^2$.  
 In addition to the measurements of $\sin^2\theta^{\rm lept}_{\rm eff}$(M$_Z$), the  measurement of the on-shell EW mixing angle  $\sin^2\theta_{W}= 1- M_W^2/M_Z^2$, and the indirect measurement of $M_W$, the CDF collaboration will be publishing the normalized $\chi^2$ weights for the set of 100 NNLO  NNPDF3.0 replicas. These weights can then be used to reduce the PDF errors in other Tevatron measurements such as the direct measurement of $M_W$.
 
   Fig.~\ref{Fig4}(b) (from ref.~\cite{pdf_error}) shows  $\chi^2$ versus   $\sin^2\theta^{\rm lept}_{\rm eff}$(M$_Z$)  for MC simulation of a  CMS like detector with  15 fb$^{-1}$ at 8 TeV. At the LHC this technique yields a much more significant reduction in the PDF  errors.  More precise high statistics  $A_{FB}(M,y)$ data place a more stringent constraints on PDFs.  As the errors in  $A_{FB}(M,y)$  become smaller at higher luminosity, the corresponding PDF errors are also reduced~\cite{pdf_error} and no longer limit the uncertainty in the extracted values of   $\sin^2\theta^{\rm lept}_{\rm eff}$(M$_Z$), the on-shell  $\sin^2\theta_{W}$  and $M_W^{indirect}$.

 \subsection{LHC and HL-LHC measurements of $\sin^2\theta_W$ and  $M_W^{indirect}$} 
 A summary of the various contribution to the errors in the CDF measurement of   $\sin^2\theta^{\rm lept}_{\rm eff}$(M$_Z$)  and $M_W^{indirect}$ is given in Table \ref{Table2}. Because $A_{FB}$ is an asymmetry, the QCD scale errors
 are small.   CDF uses the difference between the results extracted using
 LO and NLO templates as a conservative estimate of the   QCD  EBA error. This small error will become even smaller when the differences between NLO and future NNLO analyses are used as the QCD EBA error.  
The most recent (2015)  measurements of the electroweak mixing angle at the LHC
are $\sin^2\theta^{\rm lept}_{\rm eff}$(M$_Z$)=0.2398 $\pm$0.0005 (stat) $\pm$ 0.0006 (syst) $\pm$0.0009(PDF),
 reported by ATLAS\cite{ATLAS}  using 7 TeV data,  and
$\sin^2\theta^{\rm lept}_{\rm eff}$(M$_Z$) = 0.23142 $\pm$ 0.00073 (stat) $\pm$ 0.00052(sys) $\pm$ 0.00056 
(theory/PDF) reported by LHCb \cite{LHCb} using both  7 and 8 TeV data. 

The  uncertainties in the ATLAS measurement of 0.2398 $\pm$0.00120 (total)
and in the LHCb measurement of 0.23142 $\pm$0.00110 (total)  are a factor of 2 larger
than in the  corresponding measurements at the Tevatron of 0.23146 $\pm$ 0.00047 (D0),
and 0.23222 $\pm$ 0.00046 (CDF).

However if the novel techniques listed above are incorporated into the analyses,   much more precise measurements of    
$\sin^2\theta^{\rm lept}_{\rm eff}$(M$_Z$), 
$\sin^2\theta_W$ and $M_W^{indirect}$ can be extracted from Drell-Yan $A_{FB}$ data at the  LHC and HL-LHC. 
For example, the acceptance and efficiencies and sensitivity to  pileup effects cancel to first order if the event angle weighting\cite{weighting}  technique is used. The sensitivity to pileup is further reduced if only track information  is  included in  the isolation requirement for  dimuons.  When the new techniques are used,  the analysis is not limited by detector systematics. In addition,  as shown below,  the PDF errors can be significantly reduced by incorporating 
$\chi^2$ weighting in the analysis and the total error is only limited by statistics.  All of the novel techniques listed above are being incorporated in the current ongoing analyses of the CMS 8 TeV dimuon and dielectron data samples.

\begin{table}[h]
\caption { Expected statistical and $\chi^2$  weighted 
PDFs errors  in  the measurements of  $ \sin^2\theta_W$ 
and $ M_W^{indirect}$  with a CMS like detector 
for two samples. 
(1) A total of 15M reconstructed  dilepton
(8.2 M $\mu^+\mu^-$ and 6.8M  $e^+e^-$)  events,
which is similar to the existing  19 fb$^{-1}$
data sample   at 8 TeV
(with a CMS like detector).
(2) 120M reconstructed $\mu^+\mu^-$ events,
which is  the  sample expected  for 
200 fb$^{-1}$  at 13-14 TeV.  Table from ref. \cite{pdf_error}.}
    \begin{center}
\begin{tabular}{|c||c|c|c||}
\hline
CMS~like~detector&  20 $fb^{-1}$	&  $\approx$ 200 $fb^{-1}$		\\
Energy	&	8~TeV 	&  13-14~TeV		\\
data sample	&	current 	&  future		\\
\hline
Number of  		& 8.2M $\mu^+\mu^-$	& $\approx$ 120M $\mu^+\mu^-$\\
  reconstructed events &	6.8M $e^+e^-$	&\\
\hline
$\Delta \sin^2\theta_W$ CT10 PDF error & $\pm$ 0.00090 &  $\pm$ 0.00090\\
$\Delta \sin^2\theta_W$ NNPDF3.0 NNLO error &  $\pm$ 0.00050 & $\pm$ 0.00050\\ \hline
   $\Delta \sin^2\theta_W$&  &\\
  $\chi^2$ Weighted~PDF~ error  &	$\pm$ 0.00022 & $\pm$ 0.00014	\\
    $\Delta \sin^2\theta_W$  statistical~ error &	$\pm$ 0.00034& $\pm$ 0.00011 	\\\hline
     Stat+  $\chi^2$ weighted PDF~error &	$\pm$ 0.00040&$\pm$ 0.00018 	\\
 \hline
 \\
 \hline
$\Delta M_W^{indirect}$& MeV & MeV \\
 $\Delta M_W^{indirect}$ Statistical~ error  &	$\pm$17&  $\pm$5 	\\
 $\chi^2$ weighted ~PDF~ error &	$\pm$11& $\pm$7	\\ \hline
    Stat+  $\chi^2$ weighted PDF~ error &$\pm$20  & $\pm$9	\\
\hline\hline
\end{tabular}
\label{Table3}
    \end{center}
\end{table}
The LHC is a proton-proton collider and the direction of the dimuon pair is assumed to be the direction of the quark in the interaction.  The asymmetry is diluted by the probability that the antiquark carries a higher momentum fraction than the quark.  For rapidity close to zero,  the dilution is maximal and the forward-backward asymmetry is zero. Therefore the PDF error from uncertainties in the antiquark distributions are much larger at the LHC than at the Tevatron.  However, the use of $\chi^2$ weighting greatly reduces the PDF errors as shown below.
Table  \ref{Table3} from ref. \cite{pdf_error}  shows the expected statistical and   $\chi^2$ weighted 
PDF errors  in  the measurements of  $ \sin^2\theta_W$ 
and $ M_W^{indirect}$  with a CMS like detector 
for two samples. 
\begin{enumerate}
\item  A total of 15M reconstructed  dilepton
(8.2 M $\mu^+\mu^-$ and 6.8M  $e^+e^-$)  events for a CMS like detector. 
This sample is similar to the existing  CMS 19 fb$^{-1}$ data sample   at 8 TeV.
\item 120M reconstructed $\mu^+\mu^-$ events for a CMS like detector. 
This is similar to the  sample expected  for  CMS with
200 fb$^{-1}$  at 13-14 TeV.
\end{enumerate}

The PDF errors can also be reduced for other measurements (e.g. direct measurement of $M_W$) by using the same  $A_{FB}$ $\chi^2$ weighted constrained PDFs.  Any additional constraints from new high precisions measurements at the LHC  (e.g. $W$ asymmetry), can be added to the $\chi^2$ weights, thus further reducing the PDF errors in the measurements of all EW parameters. 
 \begin{figure}[h]
 \centering
\includegraphics[width=5.9in,height=3.5in]{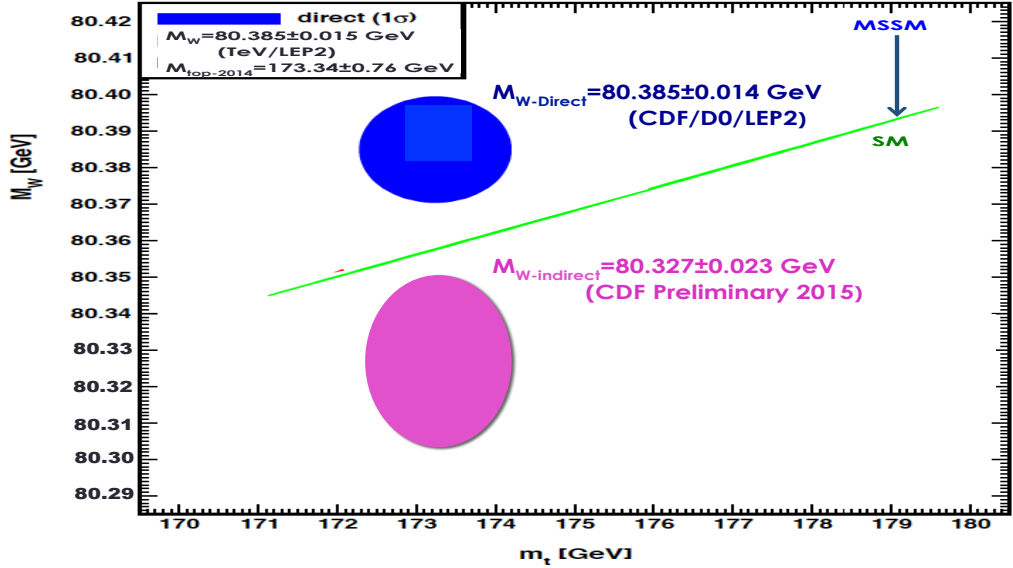}
\caption{World average of all  $M_T$ measurements  (CDF, D0, CMS, ATLAS) in  2014,
versus the most recent (2015) direct and indirect measurements of $M_W$. In the next few years
we expect a factor of 2 to 3 reduction in the uncertainties in the measurements of $M_T$, $M_W^{direct}$,
$\sin^2\theta^{\rm lept}_{\rm eff}$(M$_Z$) and
$M_W^{indirect}$ (through $\sin^2\theta_{W}= 1- M_W^2/M_Z^2$)}
\label{Fig5}
\end{figure}
\section{Conclusion}
Fig.~\ref{Fig5}   shows the  average of all  $M_T$ measurements  (CDF, D0, CMS, ATLAS) in  2014,
versus the most recent (2015) direct and indirect measurements of $M_W$.
In the future  we expect reductions of a factor of 2 to 3 in the measurement  errors of
$M_T$, the effective mixing angle  $\sin^2\theta^{\rm lept}_{\rm eff}$(M$_Z$), the  measurement of the on-shell EW mixing angle  $\sin^2\theta_{W}= 1- M_W^2/M_Z^2$, and the direct and indirect measurement of $M_W$.  Precise measurements of  $A_{FB}$(M,y) and $W$ boson asymmetry can be used to generate a set of
constrained  $\chi^2$ weighted PDFs that can be used to reduce PDF errors in the measurements of all EW parameters at the LHC.

\end{document}